# BALL LIGHTNING: MANIFESTATION OF COSMIC LITTLE BLACK HOLES

Mario Rabinowitz
Armor Research; lrainbow@stanford.edu
Redwood City, CA 94062-3922, U.S.A.                   AR/2

**Abstract.** A case is made that in encounters with the earth's atmosphere, astrophysical little black holes (LBH) can manifest themselves as the core energy source of ball lightning (BL). Relating the LBH incidence rate on earth to BL occurrence has the potential of shedding light on the distribution of LBH in the universe, and their velocities relative to the earth. Most BL features can be explained by a testable LBH model. Analyses are presented to support this model. LBH produce complex and many-faceted interactions in air directly and via their exhaust, resulting in excitation, ionization, and radiation due to processes such as gravitational and electrostatic tidal force, bremsstrahlung, pair production and annihilation, orbital electron near-capture by interaction with a charged LBH. Gravitational interaction of atmospheric atoms with LBH can result in an enhanced cross-section for polarization and ionization. An estimate for the power radiated by BL ~ Watts is in agreement with observation. An upper limit is found for the largest masses that can produce ionization and polarization excitation. It is shown that the LBH high power exhaust radiation is not prominent and its effects are consistent with observations.

## 1. Introduction

Black holes are bodies of such intense gravitational fields that even light is considered unable to leave them classically. Their masses range from $10^{-8}$ kg to billions of solar masses. The word "little" as used here refers to the radius of a black hole rather than its mass since a little black hole (LBH) the size of a neutron ($\sim 10^{-15}$ m) weighs as much as a mountain ($10^9$ ton $\approx 10^{12}$ kg). LBH with mass $\sim 10^{-3}$ kg and radius $\sim 10^{-31}$ m are found to be the most likely candidates to manifest themselves as ball lightning (BL). Prior to the awareness that LBH radiate appreciably, their presence on earth was considered highly unlikely, as LBH would devour the earth ~ million years. But with radiation evaporation of LBH, their lifetime in the earth's vicinity would be much less than the time it would take to ingest the earth. LBH would be unlikely on earth with Hawking radiation, because this radiation is devastating in all directions. The view of radiation from LBH presented by the Rabinowitz model (1999 a - c ) obviates both of the above problems since this radiation is beamed and considerably less than Hawking's (1974, 1975). In the Rabinowitz model, when LBH get so small that there would be appreciable exhaust radiation, the radially outward radiation reaction force propels them away from the earth. Cosmic LBH that come in from outer space and interact with the earth's atmosphere will be examined to ascertain if they are capable of displaying themselves as ball lightning.

Ball lightning is widely accepted, but still unexplained. A testable LBH model for BL is presented which explains most of the known features of BL. In this model, LBH produce visible light in interacting with the atmosphere. The BL core energy source is gravitationally stored energy which is emitted as beamed radiation by means of gravitational field emission. This beamed exhaust radiation tunnels out from LBH due to the field of a second body such as the earth, which lowers the LBH gravitational potential energy barrier and gives the barrier a finite width. This is similar to electric field emission of electrons from a metal by the application of an external field. By means of gravitational field emission, beamed radiation is emitted from LBH in a process differing from that of Hawking's model for radiated power, $P_{SH}$ (1974,1975), which has been undetected for 25 years. Belinski (1995), a noted authority in the field of general relativity, unequivocally concludes "the effect [Hawking radiation] does not exist."

Two LBH may get quite close for maximum tunneling radiation. Interestingly, in this limit, there is a similarity between the Rabinowitz model (1999 a - c), and what is expected from the Hawking model (1974,1975) in that the tidal forces of two LBH add together to give more radiation at their interface in Hawking's model, also producing a repulsive force. The power radiated from a LBH as derived in the Rabinowitz model (1999 a, c) is:

$$P_R \approx \left[\frac{hc^3}{4\pi GM}\right]\frac{\langle\Gamma\rangle c^3}{4GM} = \left[\frac{hc^6 \langle\Gamma\rangle}{16\pi G^2}\right]\frac{1}{M^2} \sim \frac{\langle\Gamma\rangle}{M^2}\left[3.42 \times 10^{35} W\right], \quad (1)$$

where M in kg is the mass of the LBH, and $\Gamma$ is the transmission probability. Interestingly, $P_R \propto \Gamma P_{SH}$. For the ultrarelativisitic emitted particles, the tunneling probability $e^{-2\Delta\gamma} = \Gamma$, where

$$\Delta\gamma \approx \frac{m_t}{h}\sqrt{\frac{2GM}{d}}\left\{\sqrt{b_2(b_2-d)} - \sqrt{b_1(b_1-d)} - d\ln\left[\frac{\sqrt{b_2}+\sqrt{b_2-d}}{\sqrt{b_1}+\sqrt{b_1-d}}\right]\right\} \quad (2)$$

For a black hole with center-to-center separation R from a body of mass $M_2$ (herein, the earth's mass, $6 \times 10^{24}$ kg), where $m_t$ is the mass of the emitted tunneling particle, d= $Mb_1(R-b_1)R/[M(R-b_1)R+M_2(b_1)^2]$, $b_1$ and $b_2$ are the classical turning points, and the solution applies for R >> $b_2$, when $M_2$ >> M.

We will look at several of the complex and variegated interactions produced by LBH in air directly and via their exhaust, resulting in excitation, ionization, and radiation due to gravitational and electrostatic tidal force, bremsstrahlung, pair production and annihilation, orbital electron near-capture by interaction with a charged LBH, etc. Physical processes are presented with as much clarity as possible. Thus some of the analyses need to be redone quantum mechanically and some need refinement before warranting equality signs. A byproduct of the analysis is a determination of the functional dependence of the relevant parameters related to black hole interaction scaling laws.

Most of the results in Sections 2 through 6 are derived independently of the model of black hole radiation, or whether or not there is black hole radiation. Near the LBH, exhaust radiation can augment ionization and excitation, but this complication will not be introduced at this time. Although a number of mechanisms are at work, polarization and ionization by the LBH gravitational tidal force is the major direct LBH interaction analyzed in this paper. The resulting BL recombination and de-excitation radiated power is found to be consistent with LBH that are small enough (~ 1 gram) that the exhaust levitation force is not negligible (Rabinowitz, 1999a). Maximum masses are found above which black holes cannot produce ionization or polarization excitation by gravitational tidal force.

Later sections will show that the Rabinowitz model of LBH beamed exhaust radiation has interactions in the vicinity of the earth consistent with observations. The analysis in these latter sections is relatively brief since references can be cited for the equations leading to the major conclusions. More detailed derivations are presented in the earlier sections, as analysis of the effects of the gravitational tidal force on an atomic scale are not available in the scientific literature.

## 2. LBH Gravitational Tidal Force Polarization and Ionization of the Atmosphere

2.1 DOMAIN OF VALIDITY OF NEWTONIAN POTENTIAL

The difference between the Einsteinian and Newtonian gravitational potential energy gets small at a radial distance r greater than 10 Schwarzchild radii ($R_H$, also called the horizon radius) from a black hole. This is true for all scales since

$$V = \frac{G(M)\gamma m}{r} < \frac{G(R_H c^2/2G)\gamma m}{10 R_H} = \frac{\gamma m c^2}{20}. \tag{3}$$

where $\gamma = (1 - v^2/c^2)^{-1/2}$. For velocities $v << c$, it is only necessary that V be smaller than 1/20 of the rest energy of the orbiting body. The Newtonian potential is a good approximation for $r > 10\, R_H$ for all sizes of black holes. Not only does Newton's law fail at high fields and very small distances due to general relativity, but both may fail, as does Coulomb's law, due to quantum effects such as the non-localization of the electron within its Compton wavelength, $2.42 \times 10^{-12}$ m.

2.2 GRAVITATIONAL POLARIZATION

2.2.1 *Gravitational polarizability*

The tidal gravitational force $F_T$ polarizes an atom,

$$F_T = \frac{(ze)^2 \partial}{4\pi\varepsilon a^3}, \quad (4)$$

where a is the unperturbed atomic radius, $\partial$ is the displacement relative to the electron cloud of the nucleus of mass $m_N \approx m$ the atomic mass, and $\varepsilon$ is the permittivity of free space. The factor on the right is the electrical harmonic restoring force with spring constant $K = (ze)^2/4\pi\varepsilon a^3$. The displacement produces both electric and gravitational dipole moments, of which the latter is

$$m_N \partial \approx m\partial = \alpha_p (F_T/m), \quad (5)$$

where $\alpha_p$ is the gravitational polarizability of the atom. Combining Equations (4) and (5) yields a general result independent of the form of $F_T$.

$$\alpha_p = \left(\frac{4\pi\varepsilon m^2}{(ze)^2}\right) a^3. \quad (6)$$

### 2.2.2 *Gravitational polarization excitation radius*

Induced polarization potential energy, where $\alpha_p$ is given by Equation (6), can result in perturbation excitation of an atom of radius $a + \partial \sim a$ when

$$\tfrac{1}{2}\alpha_p \left(\frac{F}{m}\right)^2 \sim \frac{\alpha_p}{2}\left(\frac{GM}{r^2}\right)^2 = \frac{1}{2}\left(\frac{4\pi\varepsilon m^2}{(ze)^2}\right) a^3 \left(\frac{GM}{r^2}\right)^2 >\sim h(\Delta f) - \frac{L_e^2}{2m_e a^2} \quad (7)$$

where $h(\Delta f)$ is an excitation energy for the frequency difference $\Delta f$ and $L_e$ is the electron orbital angular momentum of the polarized atom. Solving Equation (7) for the maximum polarization radius

$$r_p \sim \left[\frac{1}{2}\left(\frac{4\pi\varepsilon m^2}{(ze)^2}\right) a^3 \left(\frac{G^2 M^2}{h(\Delta f) - \frac{L_e^2}{2m_e a^2}}\right)\right]^{\frac{1}{4}}. \quad (8)$$

### 2.3 ENHANCED POLARIZATION AND IONIZATION CROSS-SECTIONS

The intense converging gravitational and/or electrostatic field of a LBH causes more molecules to be polarized and ionized than given by only kinetic considerations.

(Similar analysis applies for the electrostatic case.) The gravitational potential energy of a particle of mass m in the field of a LBH is

$$V = -\frac{GMm}{r} - p\left(\frac{GM}{r^2}\right) - \frac{\alpha_p}{2}\left(\frac{GM}{r^2}\right)^2, \tag{9}$$

where p is the permanent dipole moment, which is negligible for atoms but not for molecules. The binding energy of the molecules << the ionization potential, and they will be torn apart before getting in close enough for ionization. When the atomic collision frequency is low compared with the ionization rate due to tidal interaction with the gravitational field of the LBH, the ionization radius $r_i$ can be increased. This results in an enhanced ionization volume, i.e. an enhanced ionization cross-section $\sigma_E$.

To a first approximation, this problem will be treated as a simple central force problem in which angular momentum is conserved. Implications of atomic scattering, of ionization and scattering by the LBH exhaust, and of tidal force interactions will be neglected for now. These make orbital motion non-reentrant about the LBH as indicated by the Runge vector (or Runge-Lenz vector quantum mechanically). Scattering is negligible as an LBH enters the low density atmosphere from outer space and starts to produce ions around it, and even at high density, when the mean free path, $\lambda$ > the enhanced interaction radius. The interaction analysis here applies to both the sphere of ionization and to the sphere of polarization. The symbol $r_{ip}$ represents either the ionization or polarization radius depending on which case is considered.

We can make the problem one-dimensional with an effective potential energy

$$V_{eff} = V(r) + \frac{L^2}{2mr^2}, \tag{10}$$

where L is the conserved angular momentum of an atom about the LBH.

$$L = mvr_E = m(\Im kT/m)^{1/2} r_E = (\Im mkT)^{1/2} r_E, \tag{11}$$

where T is the temperature of the gas, $\Im \approx 3$ is the number of degrees of freedom of the particle, and $r_E$ is the enhanced ionization-polarization radius.

The radial velocity, $v_r = \left[2(E - V_{eff})/m\right]^{1/2} = 0$ at the closest approach to $r_{ip}$, for a particle that just grazes the original ionization-polarization sphere. Hence at $r = r_{ip}$, $V_{eff} = E = \frac{\Im}{2}kT$, where $\Im \approx 3$ is the number of degrees of freedom of the particle. Combining this with Equations (10) and (11) yields

$$V(r_{ip}) = \tfrac{3}{2}kT\left[1 - \left(r_E/r_{ip}\right)^2\right]. \tag{12}$$

Equation (12) gives us the enhanced ionization-polarization radius since $V < 0$.

$$r_E = r_{ip}\left[1 - \frac{1}{\tfrac{3}{2}kT}V(r_{ip})\right]^{1/2}. \tag{13}$$

Substituting Equation (6) into (9),

$$\begin{aligned}V &= -\left(\frac{GMm}{r}\right) - p\left(\frac{GM}{r^2}\right) - \frac{1}{2}\left(\frac{4\pi\varepsilon m^2 a^3}{(ze)^2}\right)\left(\frac{GM}{r^2}\right)^2 \\ &= -\left(\frac{GMm}{r}\right) - p\left(\frac{GM}{r^2}\right) - \left[2\pi\varepsilon\left(\frac{GMm}{ze}\right)^2\right]\frac{a^3}{r^4}\end{aligned}. \tag{14}$$

Thus $\quad r_E = r_{ip}\left\{1 + \dfrac{1}{\tfrac{3}{2}kT}\left[\dfrac{GMm}{r_{ip}} + p\left(\dfrac{GM}{r_{ip}^2}\right) + 2\pi\varepsilon\left(\dfrac{GMm}{ze}\right)^2 \dfrac{a^3}{r_{ip}^4}\right]\right\}^{1/2} \tag{15}$

From Equation (15), the enhanced ionization-polarization cross-section of the LBH is

$$\sigma_E = \pi r_{ip}^2\left\{1 + \frac{1}{\tfrac{3}{2}kT}\left[\frac{GMm}{r_{ip}} + p\left(\frac{GM}{r_{ip}^2}\right) + 2\pi\varepsilon\left(\frac{GMm}{ze}\right)^2 \frac{a^3}{r_{ip}^4}\right]\right\}. \tag{16}$$

Equations (15) and (16) are applicable if the particle mean free path $\lambda < r_E$. For convenience, this will be called the low density case. Whether the low or high density case is relevant is a function of both the density of the gas and the mass M of the LBH, since for small $r_E$, the mean free path $> r_E$ even above atmospheric pressure.

The electrostatic case is analogous. As a black hole becomes more and more charged, the Hawking radiation decreases until there is none. This is called an extreme Reissner-Nordstrom black hole which has no Hawking radiation . The ionization mechanism of this section may contribute to LBH charging while going through the atmosphere. This may also have occurred in the early more dense universe.

2.4 GRAVITATIONAL IONIZATION RADIUS

The gravitational tidal force due to an LBH of mass M acting on a body of mass m, mass density $\rho(x)$, cross sectional area $A(x)$, and length a at distance r from the center of an LBH yields a familiar expression when $\rho A$ is uniform:

$$F_T = \int_0^a \left(\frac{2GMx}{r^3}\right) \rho A \, dx = \frac{GMma}{2r^3}, \tag{17}$$

Gravitational and/or electrostatic tidal force can tear molecules apart and ionize the constituent atoms. For a perturbed atom that is free to accelerate, of mass m, radius a, atomic number z, and nuclear radius $\delta$, we have for the LBH tidal force acting between the nucleus and an outer electron of mass $m_e$:

$$F_T \sim \frac{GM}{r^3}\left[m\delta + zm_e(a+\delta)\right], \tag{18}$$

where $m \approx m_N$ the mass of the nucleus, and $r > 10\, R_H$. For ionization of an atom, $F_T$ exceeds the electrostatic force between the shielded nucleus and an outer electron.

$$\frac{GM}{r^3}\left[m\delta + zm_e(a+\delta)\right] > \frac{e[z-(z-1)]e}{4\pi\varepsilon a^2} - \frac{L_e^2}{m_e a^3}. \tag{19}$$

A simple shielding model is used here which works well for alkali metals, rather than more intricate shielding models. The unperturbed angular momentum of the electron is $[l(l+1)]^{\frac{1}{2}}\hbar$, where $l$ is the angular momentum quantum number.

A more rigorous solution would solve the Schroedinger equation to yield the outer electrons' energy levels and radial distances from the LBH, which at the first quantum number correspond to the perturbed atom's ionization potential and ionization radius. For a multi-electron atom, this would be a difficult and turbid task. So for simplicity and clarity, this is done here heuristically rather than rigorously.

Solving Equation (19), we find that the LBH gravitational tidal force produces ionization out to a radial distance

$$r_i \sim \left\{ \frac{GM[m\delta + zm_e(a+\delta)]}{\frac{e^2}{4\pi\varepsilon a^2} - \frac{L_e^2}{m_e a^3}} \right\}^{\frac{1}{3}}. \tag{20}$$

Equations (8) and (20) suggest the spheroidal shape of BL, which can become more complicated with the inclusion of other excitation sources and non-radial forces such as the LBH exhaust radiation.

2.5 HIGH DENSITY (CONFINED CASE) GRAVITATIONAL IONIZATION RADIUS

For two discrete masses $m_1$ and $m_2$ restrained with a massless bar of length a, the gravitational tidal force is

$$F_T = \frac{GMm_2}{r^2} - \frac{GMm_1}{(r+a)^2}. \tag{21}$$

When $m = m_1 = m_2$ and $r \gg a$,

$$F_T = GMm\left[\frac{2ra}{r^2(r^2 + 2ra + a^2)}\right] \approx GMm\left[\frac{2a}{r^3}\right], \tag{22}$$

which is the familiar form for $F_T$ as derived more generally in Equation (17).

At high densities, the atoms are not free to accelerate, being confined to very short mean free paths, like a constrained vertical bar with a mass m nearest the LBH and a mass $m_e$ at the other end. In this case

$$F_T = \frac{GMm}{r^2} - \frac{GMm_e}{(r+a)^2} \approx \frac{GMm}{r^2}, \tag{23}$$

since the mass m of the nucleus is so much greater than the electron mass, $m_e$. The tidal force tears molecules apart, and effectively ionizes the constitutent atoms within $r_{ic}$, the ionization radius for this case. As before,

$$\frac{GMm}{r_{ic}^2} \sim \frac{e[z-(z-1)]e}{4\pi\varepsilon a^2} - \frac{L_e^2}{m_e a^3}, \tag{24}$$

Solving Equation (24), the LBH gravitational tidal force produces ionization out to

$$r_{ic} \sim \left( \frac{GMm}{\frac{e^2}{4\pi\varepsilon a^2} - \frac{L_e^2}{m_e a^3}} \right)^{1/2} . \tag{25}$$

### 3. Ball Lightning Radiation

3.1 IONIZATION RATE

The ionization rate due to an LBH moving through the atmosphere is

$$\frac{dn_i}{dt} \sim \sigma \bar{v} n^2 + \sigma v_{BL} n^2 - \sigma_{re} \bar{v} n_i^2 - \sigma_{diff} \bar{v} n_i^2 , \tag{26}$$

where n is the number density of atoms, $\sigma$ is the ionization cross-section (enhanced or unenhanced depending on relative mean free path), $\sigma_{re}$ is the recombination cross-section, $\sigma_{diff}$ is the cross-section for diffusion out of the ionization sphere, $\bar{v}$ is the mean thermal velocity, $v_{BL}$ is the BL velocity, and $n_i$ is the number density of ions. The solution of Equation (26) is

$$n_i \sim \left[ \frac{A(Be^{Ct} - 1)}{(Be^{Ct} + 1)} \right] \underset{t \to \infty}{\approx} A = \left( \frac{\sigma(\bar{v} + v_{BL})n^2}{\sigma_t \bar{v}} \right)^{1/2} \tag{27}$$

where $B = (A + n_{io})/(A - n_{io})$, $n_{io}$ is the initial number density of ions, $\sigma_t = \sigma_{re} + \sigma_{diff}$, and $C = 2n[\sigma\sigma_t(\bar{v} + v_{BL})\bar{v}]^{1/2}$.

3.2 RECOMBINATION RADIATION

As the LBH moves through the atmosphere, its gravitational and/or electrostatic tidal force excites and ionizes air atoms around it, and carries the generated plasma along by gravitational and/or electrostatic attraction. At early times, the ionization time is short relative to the recombination time and the time for diffusion out of the ionization sphere. Entry into a LBH is difficult since the particle's deBroglie wavelength needs to be < ~ $R_H$ and because of conservation of angular momentum. In the presence of the LBH gravitational field and gradient, both the recombination and the diffusion times are longer than in free space, and $\sigma_{re}$ is reduced.

The equilibrium solution is obtained from Equation (27) as t gets large. In this limit the recombination rate is

$$R_{re} = [\sigma_{re}\overline{v}n_i^2] \sim \sigma(\overline{v} + v_{BL})n^2\left\{\left(\frac{\sigma\sigma_{re}}{(\sigma_{re} + \sigma_{diff})^2}\right)\left(\frac{\overline{v} + v_{BL}}{\overline{v}}\right)\right\}. \tag{28}$$

The radiation is hardly perceptable at first. In steady state, the electron-ion recombination radiated power/volume is

$$P_{re} = [R_{re}]V_i \sim \left[\sigma(\overline{v} + v_{BL})n^2\left\{\left(\frac{\sigma\sigma_{re}}{(\sigma_{re} + \sigma_{diff})^2}\right)\left(\frac{\overline{v} + v_{BL}}{\overline{v}}\right)\right\}\right]V_i, \tag{29}$$

where $\sigma = \pi r_E^2$ in the case where the ionization cross-section is enhanced as given in Section 2.3. For the small mean free path case, $\lambda < r_E$, where the ionization cross-section is not enhanced, $\sigma = \pi r_i^2$ for the unconfined case, or $\sigma = \pi r_{ic}^2$ for the confined case. $V_i$ is the ionization potential (15.5 eV for nitrogen), and n is the number density of air atoms. For the LBH mass range of interest ~1 gram, Equation (29) yields ~ Watts of radiated power in agreement with observation. Photons with 15.5 eV energy have a frequency higher than visible photons of a few eV with wavelengths between 4000 Å and 8000 Å. Energy degradation and other mechanisms like those of Sections 3.3 and 4 can result in visible light, thermal and de-excitation radiation.

A less detailed impulse transfer approach yields a negligible frictional power transfer $P = \frac{4\pi G^2 M^2 \rho}{v_{bh}}\ln\left(\frac{b_{max}}{b_{min}}\right)$ for light LBH, and ~ 10 W to the atmosphere for much heavier LBH with M ~ $10^{12}$ kg, $R_H$ ~ $10^{-15}$ m, $\rho_{atm}$= 1.3 kg/m$^3$ is the atmospheric mass density, and the weak logarithmic dependence of the ratio of the maximum to minimum impact parameters $\ln(b_{max}/b_{min})$ ~ 30.

3.3 POLARIZATION DE-EXCITATION RADIATION

Similarly to the recombination radiated power derived above, the polarization de-excitation radiated power/volume is

$$P_{de} = [R_{de}](h\Delta f) \sim \left[\sigma_{pE}(\overline{v} + v_{BL})n^2\left\{\left(\frac{\sigma_{pE}\sigma_{de}}{(\sigma_{de} + \sigma_{diff})^2}\right)\left(\frac{\overline{v} + v_{BL}}{\overline{v}}\right)\right\}\right](h\Delta f). \quad (30)$$

When the particle mean free path $\gtrsim r_{pE}$, then $r_{pE} = r_p\left[1 - \frac{1}{\frac{3}{2}kT}V(r_p)\right]^{1/2}$.

$$\sigma_{pE} = \pi r_p^2 \left\{1 + \frac{1}{\frac{3}{2}kT}\left[\frac{GMm}{r_p} + p\left(\frac{GM}{r_p^2}\right) + 2\pi\varepsilon\left(\frac{GMm}{ze}\right)^2 \frac{a^3}{r_p^4}\right]\right\}, \quad (31)$$

where $r_p$ is given by Equation (8). In the case when the particle mean free path $<< r_p$, $\sigma_{pE}$ is replaced by $\sigma_p = \pi r_p^2$.

Polarization of atomic energy levels by gravitational tidal force may be thought of as similar to the Stark effect where the electric field splits the energy levels of the orbiting electrons. A large separation of states of different orbital angular momentum can occur. LBH also produce a high frequency dielectric loss since atoms in elliptical orbits about the stationary LBH field see a time-varying gravitational field.

## 4. Quasi-Orbital Near-Capture Electron Radiation

A charged LBH with one to 10 $\pm$electron charges, ze, can form a superheavy atom/ion. Appreciable radiation will ensue when a quasi-orbiting particle of charge e falls in toward the LBH and is nearly captured. Energy E is radiated per particle at frequency $\approx f_o$, where the initial orbital velocity $v_o \approx 2\pi f_o r_o$. The emission of radiation near $f_o$ can account for the BL colors of low optical power:

$$E \sim \frac{z^2 e^2}{6\pi^2 \varepsilon_o c}\left(\frac{v_o}{c}\right)^2 \int_{f_{min}}^{f_{max}} \left[\frac{f^2(f_o^2 + f^2)}{(f^2 - f_o^2)^2}\right] df$$

$$= \frac{z^2 e^2}{6\pi^2 \varepsilon_o c}\left(\frac{v_o}{c}\right)^2 \left[f_o \ln\left(\frac{f - f_o}{f + f_o}\right) + \frac{f(f^2 - 2f_o^2)}{f^2 - f_o^2}\right]_{f_{min}}^{f_{max}} \quad (32)$$

## 5. Maximum Black Hole Masses

5.1 MAXIMUM MASS FOR TIDAL FORCE IONIZATION

The object of this section is to provide functional dependences so that maximum black hole masses can be calculated that will produce ionization or a given degree of polarizing excitation. The particular values obtained will depend on the case that is applicable and specifics related to the atoms in question.

### 5.1.1 *Unconfined atoms*

Solving Equation (20) for $r_i = R_H = 2GM/c^2$ yields the maximum LBH mass that can produce tidal force ionization in the case of unconfined gas atoms:

$$M_{max} <\sim \left\{ \left(\frac{c^6}{8G^2}\right) \frac{[m\delta + zm_e(a+\delta)]}{\left(\frac{e^2}{4\pi\varepsilon a^2} - \frac{L_e^2}{m_e a^3}\right)} \right\}^{\frac{1}{2}} \tag{33}$$

### 5.1.2 *Confined atoms*

In the case of confined atoms for $r_{i\,c} = R_H$, Equation (25) yields the maximum LBH mass that can produce tidal force ionization:

$$M_{max} <\sim \left(\frac{c^4}{4G}\right) \left\{ \frac{m}{\left[\left(\frac{e^2}{4\pi\varepsilon a^2} - \frac{L_e^2}{m_e a^3}\right)\right]} \right\}. \tag{34}$$

### 5.2 MAXIMUM MASS FOR TIDAL FORCE POLARIZATION EXCITATION

The smaller LBH masses are more effective in producing polarization excitation. As obtained from Equation (8) for $r_p = R_H = 2GM/c^2$, the maximum LBH mass that can produce a tidal force polarization excitation of energy $h(\Delta f)$ is

$$M_{max} <\sim \left(\frac{c^4}{4G^2}\right) \left\{ \frac{m^2 \, 2\pi\varepsilon a^3 G^2}{(ze)^2 h(\Delta f)} \right\}^{\frac{1}{2}}. \tag{35}$$

## 6. Estimate of Diffusion Time

At least three parallel competing interactions are involved in the air around a LBH: ionization of atoms inside the ionization region, recombination of electron-ion pairs, and diffusion of atoms outside the ionization region. As shown in Section 3.1, the ionization process is a coupled, non-linear process. In the limit of linear independent processes, the net ionization time would be:

$$T = \left[\frac{1}{t_{ion}} - \frac{1}{t_{recom}} - \frac{1}{t_{out}}\right]^{-1} = \frac{t_i t_r t_o}{t_r t_o - t_i t_o - t_i t_r}. \tag{36}$$

When an LBH enters the atmosphere and starts to ionize air around it, $t_r$ and $t_o >> t_i$, yielding $T \sim t_i$. At ordinary densities, the recombination rate << two-body collision frequency as three-body collisions are needed to conserve both energy and momentum for electron-ion recombination. (Interestingly, at very high densities the three-body collision rate exceeds the two-body rate.) This makes $t_r >> t_i$ until steady state when $t_i$ gets longer and eventually $t_i = t_r$. In terms of a random walk from $r > 10\, R_H$ near the LBH out to r, we can see why the escape time $t_o$ is so large. For force-free diffusion:

$$\overline{v} t_o = r = \sqrt{\eta}[\lambda] = \sqrt{\eta}[\overline{v} t_c] \tag{37}$$

where r is the net radial distance traversed, $\eta$ is the number of steps taken by an atom, $\lambda$ is the mean-free-path of a particle (atom or ion), and $t_c$ is the mean collision time. Thus the diffusion time out of the ionization region is

$$t_o = \frac{r^2}{\overline{v}\lambda} = \left(\frac{r}{\lambda}\right)^2 t_c = \left(\frac{r}{\overline{v}}\right)^2 \frac{1}{t_c}. \tag{38}$$

Using Equation (38) we can estimate the time it takes electron-ion pairs to diffuse out to $r = r_{BL}$, the characteristic radial length of ball lightning:

$$t_{BL} = \left(\frac{r_{BL}}{\overline{v}}\right)^2 \frac{1}{t_c} = \left(\frac{r_{BL}}{\lambda}\right)^2 t_c = \frac{r_{BL}^2}{\overline{v}\lambda}. \tag{39}$$

For a typical ball lightning radius $r_{BL} = 10^{-1}$ m, mean thermal velocity $\overline{v} = 6.7 \times 10^2$ m/sec, and mean-free-path $\lambda \sim 10^{-7}$ m, gives $t_{BL} > \sim 150$ sec. Gravitational interaction with the LBH increases both $t_{BL}$ and $t_{recom}$.

### 7. Beamed Little Black Hole Radiation Can Produce Levitation

The downwardly directed radiation (due to the earth below) from a $3 \times 10^{-4}$ kg (1/3 gm) LBH will act like a rocket exhaust permitting the LBH to levitate or fall slowly. We can estimate the upward force on the LBH from

$$M\frac{dv}{dt} = -c\frac{dM}{dt} - Mg \qquad (40)$$

where the exhaust leaves the LBH at near the speed of light, $c = 3 \times 10^8$ m/sec, $g = 9.8$ m/sec$^2$ is the acceleration of gravity near the earth's surface, and $dM/dt = -P_R/c^2$. For levitation $P_R \sim 10^6$ W. In one model, emission is mainly by the six kinds of neutrinos (Thorne et al., 1986) and in another almost entirely by gravitons (Argyres et al., 1998). The emitted power $P_R$, necessary to produce levitation, as well as the necessary masses and separations needed to produce this exhaust power are independent of the nature of the emitted particles.

### 8. LBH Exhaust Radiation Interactions

The interaction of LBH with matter is complex and many-faceted. Although LBH are extremely hot, it was previously shown that they produce negligible black body radiation in the visible spectrum. For example, a $3 \times 10^{-4}$ kg (1/3 gm) LBH having a temperature $T \sim 10^{27}$ K radiates only a neglible $< \sim 10^{-67}$ of its exhaust power in the visible spectrum (Rabinowitz, 1999a). Temperature is not critical here, as it is to Hawking, since the radiation does not have to be black body; and there need not be a specific energy distribution. So $\langle E \rangle \approx kT \sim 10^{23}$ eV could be used here instead of T. If the LBH exhaust radiation is almost entirely in gravitons (Argyres et al, 1998), then there is no concern that the exhaust will produce untoward effects. Classically, the exhaust may be distant from the LBH when the tunneled barrier is wide, but the results here do not depend upon this. Let us see how consequential the exhaust is in the worst case, if the exhaust is mainly composed of neutrinos, photons, electrons, positrons, and other particles. The LBH exhaust will diverge as an LBH approaches the earth. For worst case, let us focus on the undiverged LBH exhaust.

In addition to the gravitational (and equivalent electrostatic) tidal force analysis for polarization, excitation, and ionization in preceding sections, an exhaustive list of other interactions relevant to both neutral and charged particles should include:

- Bremsstrahlung
- Gravitational Scattering
- Pair Production and Showers
- Collisional Ionization
- Pair Annihilation
- Photoelectric Absorption
- Compton Scattering
- Cerenkov Radiation
- Coulomb Scattering
- Transition Radiation (Lilienfeld, 1919)
- Attenuation of Radiation in the Region of the Source (optical opacity, etc.)

Let us consider a few of these topics. Since radiation produced directly by interaction with LBH has been covered in preceding sections as well as previously (Rabinowitz, 1999a), we will next focus only on the interaction of the LBH exhaust with external matter to ascertain if the LBH beamed radiation is consistent with observations.

The ratio of energy loss per unit length by radiation and by ionization (Rasetti, 1936) for exhaust electrons with $E \sim 10^5$ to $10^{17}$ MeV is

$$[dE/dx]_{rad} / [dE/dx]_{ioniz} = zE_{Mev}/800$$
$$\sim 7(10^5 \text{ to } 10^{17})/800 \sim 10^3 \text{ to } 10^{15} \tag{41}$$

in interacting with atoms of atomic number z. Since energy loss due to ionization by the exhaust is neglible in comparison, we shall deal mainly with radiation and processes that lead to radiation.

### 9. Relativistic and Non-Relativistic Power Radiated by Charged Particles

The extremely high energy and profoundly relativistic nature of the beamed exhaust (directly emitted) radiation from LBH make it difficult to calculate its interaction in producing the visible light that is associated with BL. This is best illustrated by a simple comparison of the power radiated by an accelerated charged non-relativistic particle with the same particle in the relativistic case. Non-relativistically:

$$P_{nr} = \frac{z^2 e^2 a^2}{6\pi\varepsilon c^3}, \tag{42}$$

where ze is the total charge, a is the acceleration of the particle, ε is the permittivity of free space, and c is the speed of light. Relativistically (Rohrlich, 1965):

$$P_r = \frac{z^2 e^2 \gamma^4}{6\pi\varepsilon c^3}\left[a^2 + \left(\frac{\gamma}{c}\right)^2 (\vec{v} \cdot \vec{a})^2\right], \tag{43}$$

where v is the velocity of the particle, and $\gamma = (1 - v^2/c^2)^{-1/2}$. For particles with $E \sim 10^{23}$ eV, $E = \gamma mc^2$ implies $\gamma \sim 2 \times 10^{17}$. If $\vec{v}$ is perpendicular to $\vec{a}$, the ratio of Equation (43) to Equation (42) is $P_r/P_{nr} = \gamma^4 \sim 10^{69}$. If $\vec{v}$ ($\approx c$) is parallel to $\vec{a}$, the ratio is $P_r/P_{nr} = \gamma^6 \sim 10^{104}$. This indicates that $P_r$ can easily be overestimated or underestimated. This regime is more than 50 orders of magnitude beyond anything that has been experimentally tested.

### 10. Bremsstrahlung Radiation Due to LBH Relativistic Exhaust Electrons

The above power loss $\propto$ (acceleration)$^2 \propto (z/m)^2$ is implicit in Equation (43) and is why it occurs in the bremsstrahlung loss $\propto z^2 r_{class}^2 = z^2[e^2/4\pi\varepsilon mc^2]^2 \propto z^2$[classical radius of the charged particle]$^2$ in Equation (45). Since it is $\propto m^{-2}$, the losses are quite small for heavy particles like mesons, protons, etc. Thus we will deal primarily with electrons and positrons. Let us estimate the power radiated by those extremely high energy exhaust electrons that get close enough to the nucleus (charge ze) of target atoms to be accelerated to produce bremsstrahlung. We shall be dealing with a regime that is energetically far beyond that which is commonly dealt with.

At a distance r from the nucleus, the nuclear electric field is $E_{lab} = Ze/r^2$ as seen by the electron in the lab frame. We may equally well consider a nucleus to be moving past an electron at rest. In the electron frame, the perpendicular component of the nuclear electric field is enhanced by $\gamma$, $E_{ef} = \gamma Ze/r^2$. Thus in the electron frame, the field of the approaching nucleus looks somewhat like a plane electromagnetic wave whose pulse width is $\sim r/\gamma$.

To a first approximation, the pulse is a Gaussian of time width $r/\gamma c$, $E_{ef} \propto e^{-\frac{t^2 \gamma^2 c^2}{2r^2}}$. It follows that the amplitude distribution of component frequencies is another Gaussian of frequency width inverse to the time width, i.e. $\gamma c/r$.

Using the Thomas-Fermi model, the nucleus is screened by nearby electrons and its field is not felt at distances greater than about $a_o/z^{1/3}$, where $a_o = 0.5$Å is the Bohr radius. After arduous calculation in the electron frame, one finds by transforming to the lab frame that the cross-section per nucleus for electron scattering detected at the frequency f in the range $\Delta f$ is

$$\sigma_{brem} = 2.29 \times 10^{-31} z^2 \frac{\Delta f}{f} \ln\left(\frac{183}{z^{1/3}}\right) m^2. \tag{44}$$

So the total average energy loss per electron per m of path length (Sauter, 1934; Heitler, 1944) is

$$\left\langle \frac{dE}{dx} \right\rangle_{brem} = -\int_{f_{min}}^{f_{max}} hf\rho 2.29 \times 10^{-31} z^2 \frac{1}{f} \ln\left(\frac{183}{z^{1/3}}\right) df$$

$$= \frac{-4}{137} r_{class}^2 h(f_{max} - f_{min})\rho z^2 \ln\left(\frac{183}{z^{1/3}}\right) \tag{45}$$

$$= 2.29 \times 10^{-31} h(f_{max} - f_{min})\rho z^2 \ln\left(\frac{183}{z^{1/3}}\right)$$

$$= 5.18 \times 10^{-27} \left[\lambda_{min}^{-1} - \lambda_{max}^{-1}\right] J/m \tag{46}$$

where $\rho$ is the number of target nuclei/m³ of atomic number z. In the visible range the wavelengths are $\lambda_{max} = 8000$Å $= 8000 \times 10^{-10}$m and $\lambda_{min} = 4000$Å $= 4000 \times 10^{-10}$m; $\rho_{air} = 2.5 \times 10^{25}$/m³, gives a loss per electron of $\langle dE/dx \rangle_{air,vis} = 6.5 \times 10^{-21}$ J/m $= 4.0 \times 10^{-2}$ eV/m for its contribution to the perceptible light from BL.

In air, for all frequencies, $hf_{min} = 0$, and $hf_{max} = \gamma mc^2 = 2 \times 10^{17}(0.5 \times 10^6 eV) = 10^{23}$ eV. Thus per electron $\langle dE/dx \rangle_{air\,total} = 1.3 \times 10^{20}$ eV/m $= 13$ J/m $= 1.3 \times 10^6$ erg/cm in air. For easy comparison with just an increased density, using $z \sim 7$ in the ground with average density $\approx 2.7$ gm/cm³, $\rho_{gnd} \sim 3 \times 10^{28}$/m³ gives $\langle dE/dx \rangle_{gnd\,total} = 1.5 \times 10^{23}$ eV/m $= 2.4 \times 10^4$ J/m per electron.

The radiation range is obtained from

$$\left\langle \frac{dE}{dx} \right\rangle_{brem} = \frac{-E}{R_{rad}} \Rightarrow E = E_o e^{-x/R_{rad}}. \tag{47}$$

For $E = E_o e^{-1}$, with $E = hf$, $R_{rad}$ is obtained from Equation (45)

$$R_{rad} = \left[ 2.29 \times 10^{-31} \rho z^2 \ln\left(\frac{183}{z^{1/3}}\right) \right]^{-1}. \tag{48}$$

For air, $R_{rad\ air} = 8 \times 10^2$ m. For the ground, $R_{rad\ gnd} \approx 1$ m. Although perpendicular deflections greatly predominate over the $\vec{v}$ parallel to $\vec{a}$ case, with large $\gamma$ this case needs to be included, as it can dominate the radiated power.

## 11. Pair Production by LBH Exhaust Photons

The dominant energy loss of photon interaction with matter is by pair production. Electron-positron pair production may be thought of as the absorption of a photon in the field of a nucleus with the excitation of an electron from a negative energy state to a positive energy state. The hole left behind in the negative sea of electrons acts like an electron of opposite sign, i.e. a positron. It is like a reverse bremsstrahlung process. A simple procedure similar to the above bremsstrahlung calculation can be used to calculate the cross-section for pair production when the photon energy is much, much larger than the rest energy of the particle pair. For a photon of energy hf, let us transform to a coordinate system in which the nucleus moves toward the photon at a high velocity v < c. As seen in this new system, the photon has reduced frequency f'. As in Section 10, the nucleus looks like a plane wave of photons. By proper choice of v, the nuclear electromagnetic field can look like a wave of photons of the same reduced frequency f' as the incident photon. Then the calculation is reduced to the collision of photons in vacuum: $hf' + hf' \to e^- + e^+$.

Hence for $hf \gg 2mc^2$, the cross-section for pair production (Heitler, 1944) by the photon near a nucleus of atomic number z is found to be:

$$\sigma_{pair} = 5.72 \times 10^{-32} \left[ 3.11 \ln\left(\frac{183}{z^{1/3}}\right) - .074 \right] z^2 \text{ meter}^2 / \text{photon} \propto z^2 r_{class}^2. \tag{49}$$

Since $\sigma_{pair} \propto r_{class}^2 \propto m^{-2}$ where m is the mass of one of the pair produced particles, it is much smaller for heavy particles like mesons, protons, etc. For air, $z \approx 7$, and $\sigma_{pair,air} = 4 \times 10^{-29}$ m²/photon. For the same $z \sim 7$, we get the same photon pair production cross-section in the ground, $\sigma_{pair,gnd} \approx 4 \times 10^{-29}$ m²/photon. Electron-positron pairs thus produced will go on to radiate as calculated in Section 10.

The mean free photon path $R_{pair}$ due to pair production is obtained from the decrease in intensity of a beam of photons:

$$\frac{dn}{dx} = -\frac{n}{R_{pair}} \Rightarrow n = n_o e^{-x/R_{pair}}, \tag{50}$$

where $R_{pair} = 1/(\rho \sigma_{pair})$ for a target of number density $\rho$. Thus

$$R_{pair} = \left[ \rho 5.72 \times 10^{-32} \left\{ 3.11 \ln\left(\frac{183}{z^{1/3}}\right) - .074 \right\} z^2 \right]^{-1} \text{ meter}. \tag{51}$$

For air, $R_{pair,air} = 1.0 \times 10^3$ m. In the ground, $R_{pair,gnd} \approx 1$ m. The pairs will quickly reduce the intensity of the exhaust photons and produce their own bremsstrahlung with a concomitant shower avalanche of electron-positron pairs. Since the range is from $\sim 1$ m in the ground to $\sim 10^3$ m in air, the shower is low energy density. Annihilation by antiparticles produced by the exhaust photons as well as those originally in the exhaust result in precise high frequency gamma-rays, e.g. $e^+ + e^- \rightarrow 2hf = 1$ MeV.

## 12. Ball Lightning Incidence Rate

It was demonstrated in preceding sections that direct interaction of LBH with the atmosphere can account for the power radiated from ball lightning, that the high power LBH exhaust radiation is not prominent, and its effects on the environs are consistent with observations, let us now examine the incidence of BL on the earth. The distribution of LBH masses is not known, nor is the gravitational field enhancement of the number density of LBH near galaxies. As an upper limit calculation, let us assume that all of the dark matter, 95 % of the mass of the universe (Rabinowitz, 1999a) is uniformly distributed with 10% of the LBH mass $\overline{M}_{LBH} \sim 10^{-3}$ kg. Thus the number density of LBH is

$$\rho_{LBH} \sim \frac{10^{-1}\left[0.95 M_{univ} / \overline{M}_{LBH}\right]}{V_{univ}}. \tag{52}$$

For $M_{univ} \sim 10^{53}$ kg and $V_{univ} \sim 10^{79}$ m³ (radius of $15 \times 10^9$ light-year = $1.4 \times 10^{26}$ m),

$\rho \sim 10^{-24}$ LBH/m$^3$.

LBH whose mean relative velocities $v_{LBH}$ decrease (cf. Section 7) as they approach the earth from a distance large compared with the earth's radius due to the radiation reaction force (Rabinowitz, 1999a) are observed as BL. At large velocities, those that do not slow down appreciably due to their large mass or angle of approach, either do not produce sufficient ionization to be seen or do not spend sufficient time in the atmosphere to be observed. Recall from Section 6 that the diffusion time $> \sim 150$ sec for atoms out of the excitation region near the LBH to a BL radius of 10 cm.

Assuming no creation or loss of LBH in the time scales of interest, the continuity equation implies that the LBH current is conserved. Neglecting gravitational focussing, the LBH flux is conserved. For those LBH that manifest themselves as ball lightning, $v_{LBH}$ will decrease to the ball lightning velocity of $v_{BL} \sim 1$ m/sec., and their number density will increase as they approach the earth. In this case, the ball lightning flux is

$$\rho_{BL} v_{BL} = \rho_{LBH} v_{LBH}. \tag{53}$$

A reasonable velocity for $v_{LBH}$ is $\sim 6.2 \times 10^5$ m/sec, which is the velocity of our local group of galaxies with respect to the microwave background, i.e. with respect to the cosmic rest frame (Turner and Tyson, 1999). Since the LBH were created during the big bang, they may be expected to be in the cosmic rest frame at large distances from the earth. Equation (53) thus estimates that the ball lightning flux is $\sim 10^{-18}$/m$^2$-sec = $10^{-12}$/km$^2$-sec = $10^{-5}$/km$^2$-year without gravitational focussing. This flux implies that the incidence rate for BL $\sim 10^3$ BL/year around the world. Barry and Singer (1988) give a value of flux of $3 \times 10^{-11}$ km$^{-2}$ sec$^{-1}$, whereas Smirnov (1993) estimates $6.4 \times 10^{-8}$ km$^{-2}$ sec$^{-1}$ to $10^{-6}$ km$^{-2}$ sec$^{-1}$. In any case the BL incidence is much less than that of lightning. Even if 100 % of the dark matter were $\sim 10^{-3}$ kg LBH, an incidence rate of $10^{-11}$/km$^2$-sec is well below the noise level of present day particle detectors.

To put this in perspective, an estimate for the incidence of lightning discharges is about $3 \times 10^{-5}$/km$^2$-sec (Turman, 1977). This is why although almost everyone has seen lightning, relatively few have seen BL. Given that $v_{BL}$ ranges from $10^{-2}$ to $10^2$ m/sec with a typical value ~ 1 m/sec, Equation (53) can also yield an estimate for the number density of LBH in the region of the earth as $\rho_{eLBH} = \rho_{BL} \sim 10^{-17}/m^3$, which implies that there are ~ $10^4$ LBH here at any one time. Gravitational focussing is difficult to estimate, but could increase $\rho_{BL}$ and the BL incidence flux $\rho_{BL} v_{BL}$ by as much as a factor of $10^4$.

## 13. Discussion

Greatly decreased radiation relative to that in the Hawking model permits LBH to be prevalent throughout the universe. So it is reasonable to surmise that they are also present in the region of the earth, and may manifest themselves as BL. An interesting related question is how LBH manifest themselves in the region (accretion disk) around very large black holes where they may help initiate ionization, and provide a distinctive signature as they fall into a big black hole. In my radiation model, primeval LBH of small mass may still exist, and their beamed radiation may contribute to the accelerated expansion of the universe (Rabinowitz, 1999c).

The following criteria are presented as a guide for assessing ball lightning/earth light models in general, and the little black hole model in particular. These are derived from several sources (Fryberger, 1994; Singer, 1971; Smirnov, 1993; and Uman, 1968). LBH meet most of the criteria for BL (Rabinowitz, 1999a): 1. Constant size, brightness, and shape for times < ~10 sec. 2. Untethered high mobility. 3. Generally don't rise. 4. Can enter open or closed structures. 5. Can exist within closed conducting metal structures. 6. Levitation. 7. Low power in the visible spectrum. 8. Rarity of sightings. 9. Relatively larger activity near volcanoes. 10. Abate quietly. 11. Extinguish explosively occasionally. 12. Related radioactivity. 13. Typical absence of deleterious effects. 14. Occasional high localized energy deposition. 15. Larger activity associated with thunderstorms. 16. Bounce. ($F_{rad}$ restoring force, increases as BL goes down. The bounce frequency is related to the square root of the ratio of the restoring force to the LBH mass.) 17. Mass and Velocity Ranges. 18. Go Around Corners.

-23-Ball lightning may extinguish by different mechanisms that relate either to the LBH or to the ionized atmosphere around the LBH. One gentle mode may simply be when the sphere surrounding the LBH becomes optically opaque, or the LBH enters an opaque medium. Another more violent mode may be when this sphere blows apart because the energy dissipation in the sphere can only go out at a limited rate and becomes too great to absorb into the LBH because of conservation of angular momentum. The result is likely an explosive release of the energy. Another mode is when the LBH evaporates to such a small mass and/or loses charge so that it is not an effective ionizer of the atmosphere; the LBH is repelled away from earth by the exhaust radiation; or explodes catastrophically.

## 14. Conclusion

The visible and total radiation from the examined interactions of a LBH and its exhaust with the surrounding air are consistent with ball lightning observations. The LBH exhaust has smaller effects in the region of the earth than calculated due to divergence of the exhaust beam. There is very little directly visible exhaust radiation from LBH. Gravitational and/or electrostatic tidal interaction of a LBH with air atoms can account for the reported visible radiation. A case has accordingly been made, which deserves serious consideration, that ball lighning is the manifestation of astrophysical LBH as they interact with the earth's atmosphere. Thus the entire earth may be thought of as a laboratory for the detection of incident little black holes.

## Acknowledgment

I am grateful to Mark Davidson, Arthur Cohn, and Michael Peskin for helpful discussions.